\font\bbf       =cmbx12
\overfullrule 0pt
\vsize=9truein
\hsize=6.5truein

\def \A {{\hbar \omega^3 \over 2\pi^2 c^3}}

\def \Is {$I_*$}

\def \It	{$I_{\tau}$}

\def \Isl {$I_{*,L}$}
\def \Itl	{$I_{\tau,L}$}
\def \Itzl	{$I_{\tau=0,L}$}

\def \E	{{\bf E}}
\def \B	{{\bf B}}

\def \g  {{\bf g}}

\def \w {\omega}
\def \l {\lambda}
\def \ew {\eta(\omega)}

\def \Nzp	{{\bf N}^{zp}}
\def \Fzp	{{\bf f}^{zp}}
\def \Fszp	{{\bf f}_*^{zp}}
\def \Ezp	{{\bf E}^{zp}}
\def \Bzp	{{\bf B}^{zp}}

\def \t {\tau}

\def \ptg {$(c^2/g, 0, 0)$}
\def \r {{\bf r}}
\def \H {\left( {\hbar \omega \over 2 \pi^2} \right)^{1/2}}
\def \ekl {\hat{\epsilon}  ({\bf k}, \lambda)}
\def \kdr {{\bf k}\cdot{\bf r}}
\def \tkl {\theta ({\bf k}, \lambda)}
\def \vc {{{\bf v} \over c}}
\def \e {\hat{\epsilon}}
\def \atc  {\left( { \alpha\t \over c} \right) }
\def \tatc {\left( {2\alpha\t \over c} \right) }
\bigskip
\centerline{\bbf Gravity and the Quantum Vacuum Inertia Hypothesis}
\bigskip
\bigskip
\centerline{Alfonso Rueda}
\centerline{\it Department of Electrical Engineering}
\centerline{\it California State University, 1250 Bellflower Blvd.,
Long Beach, CA 90840}
\centerline{arueda@csulb.edu}
\bigskip
\centerline{Bernard Haisch}
\centerline{\it Chief Science Officer}
\centerline{\it ManyOne Networks, 100 Enterprise Way, Bldg. G-370, Scotts Valley, CA 95066}
\centerline{haisch@calphysics.org}
\bigskip
\centerline{(Annalen der Physik, 2005, in press)}
\bigskip
\bigskip\noindent
{\bf Abstract:} In previous work it has been shown that the electromagnetic quantum vacuum, or
electromagnetic zero-point field, makes a contribution to the inertial reaction force on an accelerated
object. We show that the result for inertial mass can be extended to passive gravitational mass. As a
consequence the weak equivalence principle, which equates inertial to passive gravitational mass, appears
to be explainable. This in turn leads to a straightforward derivation of the classical Newtonian
gravitational force. We call the inertia and gravitation connection with the vacuum fields the {\it quantum
vacuum inertia hypothesis}. To date only the electromagnetic field has been considered. It remains to
extend the hypothesis to the effects of the vacuum fields of the other interactions. We propose an
idealized experiment involving a cavity resonator which, in principle, would test the hypothesis for the
simple case in which only electromagnetic interactions are involved. This test also suggests a basis for
the free parameter $\eta(\nu)$ which we have previously defined to parametrize the interaction between
charge and the electromagnetic zero-point field contributing to the inertial mass of a particle or object.

\bigskip\noindent
\centerline{\bf 1. INTRODUCTION}
\bigskip\noindent
It appears that the
electromagnetic quantum vacuum should make a contribution to the inertial mass, $m_i$, of
a material object in the sense that at least part of the inertial force of opposition to acceleration, or
inertia reaction force, springs from the electromagnetic quantum vacuum [1,2,3].  The relevant
properties of the electromagnetic quantum vacuum were calculated in a {\it Rindler frame} (a name we
proposed for a rigid frame that performs uniformly accelerated motion)
to find a force of opposition exerted by the quantum vacuum radiation to oppose the acceleration of an
electromagnetically interacting object. We call this force associated with the quantum vacuum
radiation in accelerating reference frames the {\it Rindler frame force}. It originates in an event horizon
asymmetry for an accelerated reference frame. Owing to the symmetry and Lorentz invariance of quantum
vacuum radiation in unaccelerated reference frames, the Rindler frame force is zero in constant velocity
(inertial) frames.
Not all radiation frequencies are equally effective in exerting this opposition. The effectiveness of the
various frequencies may be parametrized by a function of the form $\eta(\omega)$ whose postulation is later
justified by means of a compelling illustration. This suggests an experimental approach to measure the
electromagnetic quantum vacuum contribution to inertial mass in a simple calculable situation. The
motivation for our interpretation comes from the discovery that the resulting force proves to be
proportional to acceleration, thus suggesting a basis for inertia of matter. [1,2,3] It thus appears that
Newton's equation of motion could be {\it derived} in this fashion from electrodynamics, and it has been
shown that the relativistic version of the equation of motion also naturally follows.

The energy of a quantum harmonic oscillator is allowed to take on only discrete values, $E_n=(n+{1 \over 2})
\hbar \omega$, implying a minimum energy of $\hbar \omega / 2$. This can be viewed as a consequence of the
Heisenberg uncertainty principle. The radiation field is quantized by associating each mode with a harmonic
oscillator [4]. This implies that there should exist a ground state of electromagnetic quantum vacuum, or
zero-point, energy. Even though this appears to be an immediate and inescapable consequence of quantum
theory, it is usually argued that such a field must be virtual, since the energy density of the field would
be expected to have cosmological effects grossly inconsistent with the observed properties of the Universe,
sometimes cited as a 120 order of magnitude discrepancy.

We set this objection temporarily aside to explore an intriguing connection between
the properties of this radiation field and one of the fundamental properties of matter, i.e.
mass. (This will, in fact, suggest an approach to resolving the discrepancy.) We do so using the techniques
of stochastic electrodynamics (SED). Alternatively, we recently have used the techniques of quantum
electrodynamics (QED) for obtaining exactly the same result [5]. Even this, though, may still be labelled a
semiclassical result in the sense that we only consider the quantum structure of the field in its general
form in the Rindler frame regardless of the details of the particles it may interact with.

If one assumes, as in SED, that the zero-point radiation field carries energy and momentum in the usual way,
and if that radiation interacts with the particles comprising matter in the usual way, it can be shown
that a law of inertia can be derived for matter comprised of electromagnetically interacting
particles that are {\it a priori} devoid of any property of mass. In other words, the
${\bf f} = m {\bf a}$ law of mechanics --- as well as its relativistic counterpart --- can be traced back
not to the existence in matter of mass (either innate or due to a Higgs mechanism), but to a purely
electromagnetic effect (and possibly analogous contributions from other vacuum fields). It can be shown that
the mass-like properties of matter reflect the energy-momentum inherent in the quantum vacuum radiation
field. We call this the {\it quantum vacuum inertia hypothesis}.

There are additional consequences that make this approach of assuming real interactions
between the electromagnetic quantum vacuum and matter appear promising. It can be shown that
the weak principle of equivalence --- the equality of inertial and gravitational mass
\footnote{$^1$}{When we write {\it gravitational mass} in this paper we mean {\it passive gravitational
mass} unless otherwise specified. Only for point particles or for spherically symmetric stationary matter
distributions can passive and active gravitational mass densities be deemed identical. In the general case
they are not the same. This difference mainly appears in cosmology, where passive and active gravitional
mass densities in general differ [6].}
--- naturally follows. In the quantum vacuum inertia hypothesis, inertial and
gravitational mass are not merely equal, they prove to be the identical thing. Inertial mass arises upon
acceleration through the electromagnetic quantum vacuum, whereas gravitational mass --- as manifest in
weight --- results from what may in a limited sense be viewed as acceleration of the electromagnetic
quantum vacuum past a fixed object. The latter case occurs when an object is held fixed in a gravitational
field and the quantum vacuum radiation associated with the freely-falling frame instantaneously comoving
with the object follows curved geodesics as prescribed by general relativity.

Finally, the interactions of the quantum vacuum radiation field with massless particles
results in Schr\"odinger's zitterbewegung, consisting of random speed-of-light
fluctuations for a particle such as the electron. These fluctuations can be shown to cause a
point-like particle to appear spread out in volume over a region --- call it the Compton
sphere --- reminiscent of that predicted by the corresponding wave function. When this
is viewed from a moving reference frame, the Doppler shift of these fluctuations results in an
observed superimposed envelope that has properties like the de Broglie wave [7]. We thus tentatively
suggest that it may be worth considering the so-called rest mass of particles to be a manifestation of
the energy associated with zitterbewegung, especially since
simulations are showing that in the presence of an external field, massless particles
undergoing zitterbewegung will acquire helical, spin-like properties (L. J. Nickisch, priv. comm.).

It is thus suggestive that both mass and the wave nature of matter can be traced back
to specific interactions with the electromagnetic zero-point field and possibly the other bosonic vacuum
fields. Given this possible reinterpretation of fundamental properties, we suggest that it is premature
to take a firm stand against the reality of the zero-point radiation field and its
associated energy on the basis of cosmological arguments, especially given the possible
relation between quantum vacuum, or zero-point, radiation and dark energy.

SED is a theory that includes the effects of the electromagnetic
quantum vacuum in physics by adding to ordinary Lorentzian classical electrodynamics a
random fluctuating electromagnetic background constrained to be homogeneous and
isotropic and to look exactly the same in every Lorentz inertial frame of
reference [8,9,10]. This replaces the zero homogeneous background
of ordinary classical electrodynamics. It is essential that this background not
change the laws of physics when exchanging one inertial reference system for another.
This translates into the requirement that this random electromagnetic background
must have a Lorentz invariant energy density spectrum. The only random electromagnetic
background with this property is one whose spectral energy density,
$\rho(\w)$, is proportional to the cube of the frequency, $\rho(\w) d\w
\sim \w^3 d\w$. This is the case if the energy per mode is $\hbar
\w/2$ where
$\w$ is the angular frequency, the other factor of $\w^2$ coming from the density of modes [4].
(The $\hbar \w/2$ energy per mode is of course also the minimum energy of the analog of an
electromagnetic field mode: a harmonic oscillator.) The spectral energy
density required for Lorentz invariance is thus identical to the spectral energy
density of the zero-point field of ordinary quantum theory. For most purposes,
including the present one, the zero-point field of SED may be identified with the
electromagnetic quantum vacuum [10]. However SED is essentially a classical theory since
it presupposes only ordinary classical electrodynamics and hence SED also presupposes
special relativity (SR). Our reliance on SED [1,2,3] has been due to the fact that SED allows a very
clear intuitive representation. Nevertheless, the SED-based results of [1,2] have been readily
rederived within a QED formalism [5].

In $\S 2$ we briefly review a previously-derived development of the quantum vacuum inertia
hypothesis.
In $\S 3$ we introduce an example that allows a very clear visualization of the coupling of an
electromagnetically interacting object with the surrounding vacuum fields and the evaluation of the
function $\eta(\omega)$ mentioned above [1,2]. We turn next to the principle of equivalence. According to
the weak equivalence principle (WEP) of Newton and Galileo, inertial mass is equal to gravitational mass,
$m_i=m_g$. If the quantum vacuum inertia hypothesis is correct, a very similar mechanism involving the
quantum vacuum should also account for gravitational mass. This novel result, restricted for the time being
to the electromagnetic vacuum component, is precisely what we show in
$\S 4$ by means of formal but simple and straightforward arguments requiring physical
assumptions that are uncontroversial and widely accepted in theoretical physics.
In $\S 5$ the consistency of this argument with so-called metric theories of gravity
(i.e. those theories characterized by spacetime curvature) is exhibited. In addition
to the metric theory, {\it par excellance}, Einstein's GR, there is the Brans-Dicke
theory and other less well known ones, briefly discussed by Will [11].
A non-spacetime curvature theory is briefly discussed in $\S 6$.

Nothing in our approach points to any new discriminants among the various metric
theories. Nevertheless, our quantum vacuum approach to gravitational mass will be
shown to be entirely consistent with the standard version of GR. Next, in $\S 7$, we
take advantage of geometrical symmetries and present a short argument supported by standard
potential theory to show that in the weak field limit a Newtonian inverse square
force must result.

A new perspective on the origin of weight is presented in $\S 8$.
In $\S 9$ we discuss an energetics aspect, related to the derivation presented
herein, and resolve an apparent paradox. 
A brief discussion on the nature of the
gravitational field follows in
$\S 10$, but we infer that the present development of our approach does not provide any
deeper or more fundamental insight than GR itself into the ability of matter to bend
spacetime. We present conclusions in $\S 11$.

\bigskip\noindent
\centerline{\bf 2. THE RINDLER FRAME FORCE AND INERTIA}
\bigskip\noindent
It has been shown [1,2,3] that the
electromagnetic quantum vacuum makes a contribution to the inertial mass, $m_i$, of a material object in
the sense that at least part of the inertial force of opposition to acceleration, or
inertia reaction force, springs from the electromagnetic quantum vacuum. Specifically, in the
previously cited work, the properties of the electromagnetic quantum vacuum as experienced in a
Rindler, or constant proper acceleration, frame were investigated, and we calculated a resulting force of
opposition on material objects fixed in such a Rindler non-inertial frame and interacting with the
electromagnetic zero-point radiation field. It
originates in an event horizon asymmetry for an accelerated reference frame. Owing to the symmetry and
Lorentz invariance of zero-point radiation in unaccelerated reference frames, the Rindler frame force
vanishes at constant velocity, i.e. in inertial frames. We further hypothesized that if $V_0$ is the proper
volume of the material object, some frequency-dependent fraction,
$\eta(\omega)$, of this radiative momentum will, via interactions, be conferred upon the
particles comprising the matter. The motivation for
this interpretation comes from the discovery that the resulting Rindler frame force proves to be
proportional to acceleration. It thus appears that Newton's equation of motion can be {\it derived} in this
fashion from electrodynamics, and it has been shown that the relativistic version of the equation of motion
also naturally follows.
We have called the notion that at least part of the inertia of an object should be due to the
individual and collective interaction of its quarks and electrons with the
quantum vacuum as the {\it quantum vacuum inertia hypothesis} with the proviso that analogous contributions
are expected from the other bosonic vacuum fields.

The electromagnetic field inside a cavity of perfectly reflecting walls can be shown to be expandable in
countably infinite different modes where each mode corresponds to an independent oscillation which always,
even in the case of free space, can be represented as a  harmonic oscillator [12].
When quantized, the harmonic oscillator gives an energy of the form $\hbar \omega (n+1/2)$, where $\omega$
is the characteristic frequency of the mode $(\omega=2 \pi \nu)$, and when the integer $n$ is set to zero,
there is still a remnant energy $\hbar \omega/2$. Alternatively we also have that for the harmonic
oscillator, the Heisenberg uncertainty relation tells us that the minimum (zero-point) energy is
$\hbar \w/2$. The electromagnetic field therefore has a minimum quantum energy state consisting of
zero-point fluctuations having an average energy per mode of $\hbar\w/2$. The density of modes is, see e.g.
[4],

$$N(\w) d\w = {\w^2 d\w \over \pi^2 c^3} , \eqno(1)
$$

\noindent
and this combined with the average energy per mode of $\hbar \w /2$ yields a spectral energy density for
the zero-point fluctuations of

$$\rho(\w) d\w = {\hbar \w^3 \over 2 \pi^2 c^3} d\w . \eqno(2)
$$

In the semi-classical approach known as stochastic
electrodynamics (SED) the quantum fluctuations of the electric and magnetic fields are treated as random plane waves
summed over all possible modes with each mode having this $\hbar\w/2$ energy. The electric and magnetic zero-point
fluctuations in the SED approximation are thus

$$\Ezp(\r,t)=\sum_{\l=1,2} \int d^3 k \H \ekl \cos \left[ \kdr-\w t - \tkl \right] , \eqno(3a)
$$

$$\Bzp(\r,t)=\sum_{\l=1,2} \int d^3 k \H \left[ \hat{k} \times \ekl \right] \cos \left[ \kdr-\w t - \tkl \right] ,
\eqno(3a)
$$
\noindent
where the sum is over the two polarization states, $\e$ is a unit vector, and $\tkl$ is a random variable
uniformly distributed in the interval $(0, 2\pi)$. For simplicity and to facilitate comparison with much
previous SED work, we omit the Ibison and Haisch modification of eqns. (3) in which the amplitudes are
also randomized in such a way as to bring the quantum and classical statistics of the electromagnetic
zero-point field into exact agreement. [13]

Since the $\Ezp$ and $\Bzp$ field fluctuations are entirely random, there is no net energy-momentum flow
across any surface, or in other words, the value of the stochastically averaged Poynting vector must be
zero:

$$<\Nzp>={c \over 4\pi} \left< \Ezp \times \Bzp \right> = 0 . \eqno(4)
$$ 

It is nowadays well known how to transform the $\Ezp$ and $\Bzp$ field vectors from a stationary frame
to one undergoing uniform acceleration, i.e. having constant proper acceleration, that we have
labelled a Rindler frame. Such a frame will experience an asymmetric event horizon
leading to a non-zero electromagnetic energy and momentum flux as calculated by a stationary observer. The
velocity and the Lorentz factor for such a frame are,

$${v \over c} = \tanh \atc , \eqno(5a)
$$

$$\gamma_{\tau} = \cosh \atc , \eqno(5b)
$$
\noindent
where $\alpha$ is the object's proper acceleration and $\tau$ its proper time.

The general, compact form of the Lorentz transformation of electromagnetic fields is (cf. eqn. 11.149 in
[14])

$$\E' = \gamma (\E + \vc \times \B) - \left( {\gamma^2 \over \gamma+1} \right) {{\bf v} \over
c} (\vc
\cdot \E ) , \eqno(6a)
$$
$$\B' = \gamma (\B - \vc \times \E) - \left( {\gamma^2 \over \gamma+1} \right) {{\bf v} \over
c} (\vc
\cdot \B ) . \eqno(6b)
$$

Particularizing {\bf v} to be along the $x$-direction and using component form of eqns. (6), one
transforms $\Ezp$ and $\Bzp$ to the Rindler frame (in a way explained in considerably more detail in
[1]) in order to obtain:

$$
\Ezp(0,\t) = \sum_{\l=1,2} \int d^3 k \H \Big\{ \hat{x} \e_x +  
\hat{y} \cosh \atc \left[ \e_y - \tanh \atc (\hat{k} \times \e)_z \right]  $$
$$ 
+\hat{z} \cosh \atc \left[ \e_z + \tanh \atc (\hat{k} \times \e)_y \right]
\Big\}
\cos\left[ k_x{c^2 \over a} \cosh \atc - {\w c \over a}\sinh \atc -\tkl \right] , \eqno(7a)
$$

$$
\Bzp(0,\t) = \sum_{\l=1,2} \int d^3 k \H \Big\{ \hat{x} (\hat{k} \times \e)_x +  
\hat{y} \cosh \atc \left[ (\hat{k} \times \e)_y + \tanh \atc \e_z \right]  $$
$$ 
+\hat{z} \cosh \atc \left[ (\hat{k} \times \e)_z - \tanh \atc \e_y \right]
\Big\}
\cos\left[ k_x{c^2 \over a} \cosh \atc - {\w c \over a}\sinh \atc -\tkl \right] , \eqno(7b)
$$
where $\e_x$ signifies the scalar projection of the $\e$ unit vector along the $x-$direction, and similarly
for $\e_y$ and $\e_z$.

Our goal here has been to calculate the rate of change of the momentum applied by the zero-point field on
the electromagnetically interacting, accelerating object. Each individual inertial frame has associated with
itself its own (or proper) random electromagnetic zero-point field background and thus its
proper zero-point field vacuum spectral energy density distribution. For each inertial frame its proper
electromagnetic vacuum is homogeneously and isotropically distributed. This means that the components of
the net Poynting vector of the zero-point field of a given frame when observed in that same frame all should
vanish. Moreover, of the components of the $4 \times 4$ electromagnetic energy-momentum stress tensor only
the diagonal components survive as all other components should be zero. Extreme care should be
exercised when performing such Lorentz transformations. Each of the components of the tensor is represented
by an improper integral over
$k-$space, which is the $k-$space representation associated with the particular inertial frame that we
are dealing with. When transforming to another inertial frame it is not enough to transform the interior
of the integral, i.e. its integrand. One must also take care of what in Appendix C of [1] we called the
$k-$sphere. This is done by means of a cutoff convergence factor $e^{-k/k_c}$ that effectively cuts off
the zero-point radiation beyond some maximal magnitude of the wave vector in the $k-$space of the
original inertial frame, thus the name $k-$sphere. The cutoff parameter $k_c$ clearly makes the
zero-point energy as well as all other components of the tensor, like the Poynting vector components and
the Maxwell $3 \times 3$ stress tensor components, all finite, which is a desirable consequence.
Nevertheless, the spectral Lorentz invariance holds even when at the end we make $k_c \rightarrow \infty$
anyway. In such cases $e^{-k/k_c}$ becomes just a form factor to regularize the integration [1].

Observe however that in the many integrations we performed in [1] we really did not need to use such a
cutoff factor. The reason is clear: the final integrations were performed at accelerating object proper time
$\tau=0$, when the laboratory frame and the accelerated-particle frame instantaneously comoved. Indeed, we
observe that all final integrations for the inertia reaction force in [1] are performed at $\tau=0$,
which is the time instant when the uniformly-accelerated object comoves with the inertial laboratory
frame. This means that both the momentum density, and thereby the Poynting vector, of the zero-point
field are for the object, those of the laboratory frame.

But there is a more subtle change that needs to be captured. Both the momentum density and the total
momentum of the zero-point field inside the volume $V_0$ of the body display a time rate of change and
have a non-vanishing time derivative. Thus although both the Poynting vector and the momentum of the
zero-point field inside the body instantaneously vanish, their time derivatives at that coincidence time
do not vanish. This can also be visualized as a manifestation of a reduction in symmetry of the event
horizon that passes from the perfect three-dimensional symmetry of a sphere to a lower two-dimensional
symmetry that is merely axial. This results in a special situation that may be visualized as a stress or
tension in the vacuum field which is manifested on the accelerated particle as the Rindler frame force.

From the electromagnetic zero-point field in the Rindler frame of the accelerated object we calculate the
Poynting vector (see [1] for details). For uniform acceleration in the $x$-direction, the
Poynting vector is

$$
\Nzp(\tau) = -\hat{x} {2c \over 3} \sinh \tatc \int {\hbar \w^3 \over 2 \pi^2 c^3} d\w =  -\hat{x} {2c \over 3} \sinh
\tatc
\int \rho(\w) d\w ,
\eqno(8)
$$

\noindent
where $\rho(\w)$ is the spectral energy density of the zero-point fluctuations. 

The amount of radiative momentum carried by the zero-point fluctuations that are passing through and
instantaneously contained in the accelerated object which is undergoing uniform proper acceleration
$\vec{\alpha}=\alpha \hat{x}$ and that has proper volume
$V_0$, is

$${\bf p}^{zp} = -\hat{x} {V_0 \over \gamma_{\tau}} {2 \over 3 c} \sinh \tatc \int \rho(\w)
d\w  = -\hat{x} {4 \over 3} V_0 v_{\tau} \gamma_{\tau} \left[ {1 \over c^2} \int \rho(\w) d\w
\right]. \eqno(9)
$$

\noindent
Let us assume that at frequency $\omega$ some small fraction
$\eta(\w)$ of this energy interacts with the particles of matter contained in
$V_0$. We can then write the interacting fraction of the zero-point momentum instantaneously contained in
and passing through the object as

$${\bf p}^{zp} = -\hat{x} {4 \over 3} V_0 v_{\tau} \gamma_{\tau} \left[ {1 \over c^2} \int
\eta(\w) \rho(\w) d\w \right] .
\eqno(10)
$$

\noindent
If we take the time derivative of this quantity, we find that [1]

$${d{\bf p}^{zp} \over dt} = \Fzp = - \left[ {4 \over 3}{V_0
\over c^2}
\int
\eta(\w)
\rho(\w) d\w \right] {\bf a}.
\eqno(11)$$

\noindent
A fully covariant analysis eliminates the factor of 4/3 and also yields a proper relativistic four-vector force
expression (see Appendix D of [1]). Eqn. (11) is telling us that in order to maintain the acceleration of
such an object, a motive force, {\bf f}, must continuously be applied to balance the vacuum electromagnetic
counteracting reaction force, that we may call the Rindler frame force, $\Fzp$. The motive force is,

$${\bf f} = -\Fzp= \left[ {V_0 \over c^2} \int \eta(\w) \rho(\w) d\w \right] {\bf a} \ . \eqno(12)
$$

\noindent
where we
have suppressed the factor of 4/3. This strongly suggests Newton's ${\bf f} = m {\bf a}$, and that the
zero-point field contribution to the inertial mass contained in
$V_0$ is the mass-like coefficient

$$
m_i=\left[ {V_0 \over c^2} \int \eta(\w) \rho(\w) d\w \right] . \eqno(13)
$$
\noindent
In other words, some of the apparent inertial mass of an object originates in the interacting
fraction of the zero-point energy instantaneously contained in an object. In this view, the apparent
momentum of the object can be traced back to the momentum of the zero-point radiation field.

We suspect that
$\eta(\w)$ involves some kind of resonance at the Compton frequency of individual particles, $(\w_C= m c^2 /\hbar)$,
since this suggests a close connection between the origin of mass and the de Broglie wavelength, both stemming from
interactions of matter with the quantum vacuum [7]. 

We speculatively digress for a moment in order to try to gain some
physical insight into how the interrelationships of minutely different coordinate systems can translate
into real measurable effects. Since momentum is not absolute but relative to the observer, and indeed always
zero in one's own frame, one wonders how the time derivative of this arbitrary quantity can yield something
real and absolute, i.e. a force. We suggest that the answer is deeply rooted in the basis of special
relativity that all interactions are limited by the speed of light. Hence if one pushes on an object at
point A, it will take a finite time for that push to be transferred to point B in the same object. The
acceleration of any object of finite extent involves a co-mingling of an infinite number of minutely
different reference frames.
Acceleration necessarily mixes infinitely many adjacent reference frames as a result of the propagation
limitations of special relativity.
This analogously suggest a physical meaning for the otherwise seemingly purely
mathematical operation of calculating how
$\Ezp$ and
$\Bzp$ would look to an observer in one (inertial) reference frame from the basis of another (accelerating).

In Appendix B of [1] we also showed an alternative but complementary approach. Instead of considering the
opposition to acceleration due to the electromagnetic zero-point field background through which the object
is being accelerated (of the previous approach in the body of Ref. [1]) we also looked at the amount of
electromagnetic energy contained within the object according to the viewpoint of an inertial stationary
observer (Appendix B of [1]). The very natural result that ensues for that stationary observer is that as
the object moves faster and faster the amount of enclosed energy of zero-point radiation grows in exact
proportion to
$\gamma_{\tau}$, which is exactly the same way as the mass of a moving and accelerating object of rest mass
$m_i$ behaves according to an inertial stationary observer. In this complementary approach the rest mass
turns out to be exactly the same as in the zero-point field drag approach of eqn. (13) above. Moreover,
whereas in eqn. (10) for the zero-point field drag approach we obtained the force $\Fzp$ that the
zero-point field exerts on the object when opposing its acceleration, in the new approach we directly get
$\bf f$, the force that the external agent has to apply to the object to accelerate it, which we already
noticed in eqn. (12).

The formal development, i.e. the equations in the derivation, of this complementary approach comes out to
be supeficially similar to the formal development of the zero-point field drag approach, except that all
equations since the very first ones [1] have the opposite sign than in the previous case. Conceptually
though, the two approaches are not similar. Instead they are very complementary. We could say that
they are like the two sides of the same coin. One cannot exist without the other, but they are distinctly
different.

The rest mass in eqn. (13) exactly corresponds to the amount of zero-point field mass equivalent enclosed
within the volume of the body and that, thanks to the $\eta(\omega)$ spectral factor interacts with the body. If
the mass is moving with respect to a stationary observer, with speed $v_{\tau}$, the associated momentum is {\bf
p} and in the second approach comes as given by eqn, (11) but with the positive sign. (Recall that the 4/3
factor comes from not using a covariant approach. The covariant approach of Appendix D [1] removes the
spurious factor by considering all components in the electromagnetic stress-energy momentum tensor.)

This is analogous
to the situation of relativistic mass increase. A stationary observer would come to the conclusion that the mass
of an accelerating object is steadily increasing as $\gamma_{\tau}m$, but in the accelerating frame
no mass change is evident. One's momentum in one's own reference frame is always zero.
Physical consequences only ensue when there is a change in momentum, akin to a ``physical'' change
from one reference frame into another.

In the approach of the quantum vacuum inertia hypothesis, it becomes physically evident how the Lorentz
factor
$\gamma_{\tau}$, which characterizes a space-time geometry relationship, acquires the physical role of
relativistic mass increase parameter. {\it It also becomes clear why the relativistic mass increase must
become infinite at
$c$: One cannot propagate the effect of any forces due to the zero-point field at speeds faster than $c$.}

\bigskip\noindent
\centerline{\bf 3. ON THE ELECTROMAGNETIC MODEL OF THE ACCELERATED OBJECT}
\bigskip\noindent
The second or complementary approach of [1], Appendix B, that we mentioned above, very clearly shows that
$m_i c^2$ is just the amount of zero-point energy located inside the accelerated object and
that
instantaneously comoves with the object as viewed by
an inertial observer.

Here we would like to visualize the meaning of the parameter $V_0$, the volume of the object in an
electromagnetic sense, and the ``efficiency'' or ``interaction'' function $\eta(\w)$. For this we just need
to think of an electromagnetic cavity comprised of surrounding conducting walls. In such a case
$V_0$ is just the internal volume of the cavity. The most familiar shape is a rectagular parallelepiped of
sides $a$, $b$ and $c$, with $a < b < c$. We consider here only the volume inside the cavity and as far as
$m_i$ goes, the mass is just $1/c^2$ times the total amount of zero-point field energy that can be
accumulated within the structure of the cavity.

The cavity walls are conducting and as such serve to confine the electromagnetic oscillating radiation
within its boundaries. Internally, the cavity can sustain a countable and finite number of characteristic
or proper modes of oscillation. It is well known that beyond a threshold frequency $\w_p$ that corresponds
to the plasma frequency of the electrons in the conducting walls, microwave cavities become essentially
transparent and cannot sustain any more modes internally. The lowest frequency modes are comparatively
widely distributed in frequency but as the characteristic frequencies of the modes increase, they
become more closely spaced in frequency. For wavelengths $\l$ of the modes much smaller than the dimensions
of the cavity, $\l << a < b < c$, their frequency number density starts to grow roughly in proportion to
$\w^2 / \pi^2 c^3$ which, not surprisingly, is the $N(\w)$ of eqn. (1). Recall that $N(\w)$ is obtained
when the cavity that encloses the radiation grows in size by letting $c > b > a \rightarrow \infty$, which
yields what we call the {\it limit of the continuum}.

The walls, because they are conducting, serve to indirectly connect the inside of the cavity to the
outside. At zero temperature, on the outside we have the zero-point electromagnetic radiation of eqns. (1)
and (2) and in the inside only the frequencies corresponding to the modes which can be excited. The modes of
the cavity weakly couple through the conducting walls to the random electromagnetic radiation outside. At
zero temperature, the cavity $l^{th}$ mode, being a harmonic oscillator, acquires the oscillation of
energy $\hbar
\w_l /2$ where $\w_l$ represents the $l^{th}$ mode characteristic frequency.

Let us first assume that the cavity is an ideal one in which dissipation may be neglected. This by itself
does not preclude some broadening around the characteristic frequency because there are, in the walls,
quantum fluctuations of the electrons and of the plasma of electrons at zero temperature which induce some
small albeit nonvanishing blurring of the exact mode characteristic frequency due to the Doppler shifts
and fluctuations in the cavity geometry (dimensions $a$, $b$ and $c$ are not exact anymore). Hence
there is always some broadening.

For the time being, however, let us neglect such broadening and consider instead an ideal cavity with a
finite number of discrete modes up to $\w_p$, the plasma frequency of the electrons in the walls, and at
strictly zero temperature, $T=0$. Each mode then oscillates at its exact characteristic frequency, $\w_k$,
and with an energy corresponding to that of the harmonic oscillator zero-point oscillation, $\hbar \w_k
/2$, \ $k=1, \ 2, ... , N$, where $N$ is the maximal mode frequency such that $\w_N \le \w_p$. The total
energy is then
$$E= \sum_{k=1}^N {\hbar \w_k \over 2}  \ , \ \ \ \  \w_1 \le \w_2 \le \w_3 \le ... \ \w_N \le \w_p \ .
\eqno(14)
$$
But this energy must be exactly equal to the energy given by eqn. (13):
$$m_i c^2 = \int_{\w=0}^\infty V_0 \ \eta(\w) \rho(\w) d\w = E = \sum_{k=1}^N {\hbar \w_k \over 2} \ .
\eqno(15)
$$
As $\rho(\w)$ is given by eqn. (2), it is easy to see that in order to satisfy (15) we need an $\eta(\w)$
of the form:
$$\eta(\w) = \sum_{k=1}^N {1 \over V_0} \cdot {\pi^2 c^3 \over \w^2} \delta(\w-\w_k)
= \sum_{k=1}^N {\pi^2 c^3 \over V_0} {1 \over \w^2} \delta(\w-\w_k) \ . \eqno(16)
$$

The idealized cavity that we have considered gives rise to an $\eta(\w)$ that essentially consists of a
finite sum of idealized line-shaped functions of the usual form present when dissipation is neglected.

A better model is obtained when we allow for some dissipation in the interaction of the modes with the
walls, or if no dissipation is considered, some broadening must still be present because of the reasons
indicated above. It is well-known that in many cases, if the dissipation is small (see, e.g.
[15,16,17]), that the lineshape function is the so-called Lorentzian lineshape function
$$g(\w)d\w = {\Delta \w \over \displaystyle 2\pi \left[ (\w - \w_0)^2 + \left( {\Delta \w \over 2} \right)^2
\right]} \ d\w \ . \eqno(17)
$$

The ubiquity of the Lorentzian lineshape comes from the fact that it originates in a dissipative
exponential time-dependent decay (see [18]). The function $g(\w)$ is called the Lorentzian lineshape
function and it is such that
$$\int_0^\infty g(\w) d \w = 1 \eqno(18)
$$
and that
$$\lim_{\Delta \w \rightarrow 0} g(\w) = \delta (\w - \w_0) \ . \eqno(19)
$$

The lineshape broadening parameter $\Delta \w$ is due to various forms of dissipation and other broadening
effects --- such as those mentioned above --- and $4\pi / \Delta \w$ is the exponential time constant of
decay for the relevant dissipative part of the process.

In the more realistic case when line broadening around the modes characteristic frequencies is considered,
combining the inputs from (16) and (17) yields
$${\cal K} (\w) \equiv V_0 \eta(\w) = {\pi^2 c^3 \over \w^2} \sum_{k=1}^N
{\Delta \w_k \over \displaystyle 2\pi \left[ (\w - \w_k)^2 + \left( {\Delta \w_k \over 2} \right)^2
\right]} \eqno(20)
$$
where, as explained above, the $\w_k$ is the characteristic frequency of the $k^{th}$ mode and $\Delta
\w_k$, with $\Delta \w_k >0$, its corresponding line broadening. Only at the highest frequencies in the
microwave cavity are the modes expected to overlap because the $\Delta \w_k$ broadenings become comparable
to the frequency separations between the modes.

All in all the corresponding contribution of eqn. (20) to the inertial mass of a physical cavity resonator
when the walls are included is fairly small, but what is in principle interesting is that it is not
negligible! The inertial mass associated with the cavity itself (without the mass of the walls) of eqn.
(20) becomes
$$\eqalignno{ m_i &= {1 \over c^2} \int_{\w=0}^\infty V_0 \ \eta(\w) \rho(\w) d\w \cr
&= \int_{\w=0}^\infty {\pi^2 c \over \w^2} \sum_{k=1}^N {\Delta \w_k \over \displaystyle 2\pi \left[ (\w -
\w_k)^2 + \left( {\Delta \w_k \over 2} \right)^2
\right]} \ \cdot \ {\hbar \w^3 \over 2 \pi^2 c^3} d\w \cr
&={1 \over c^2} \sum_{k=1}^N \int_{\w=0}^\infty {\Delta \w_k \over \displaystyle 2\pi \left[ (\w - \w_k)^2
+ \left( {\Delta \w_k \over 2} \right)^2
\right]} \ \cdot \ {\hbar \w \over 2} d\w &(21)
}
$$
which is a rather interesting expression. We clearly just add up the zero-point energies of all the
resonant cavity modes broadened by their corresponding Lorentzian broadening factors.

Observe that in eqn. (21) the volume $V_0$ of the cavity seemingly has disappeared. Of course that is just
a superficial remark as the volume as well as the geometry are determining factors in the spectrum of
frequencies $\{ \w_k \}$ distribution up to the plasma frequency of the cavity, $\w_p$, with $\w_N \le
\w_p$. The advantage for the general case of a more general object (or particle) is that $V_0$ is a
parameter that in eqn. (21) has disappeared and been replaced by electromagnetic quantities like the $\w_k$
and $\Delta \w_k$ that, among other things, depend on the geometry.

In the case of a microwave cavity resonator mode with good conducting walls, relatively, the decay times of
the modes are very long and consequently the $\Delta \w_k$ parameters are very small with $\Delta \w_k <<
\w_k$. This allows us to propose a simple experiment to measure the electromagnetic quantum vacuum
contribution to inertial mass.

If the $\Delta \w_k$ are indeed negligible for a given conductor for all modes $\w_k$ up to the maximal one
$\w_N$, where $\w_N \le \w_p$, and where $\w_p$ is the plasma frequency of the conduction-band electrons in
the metal of the walls, then the contribution to inertial mass given by the cavity is as given by eqn. (15),
a quantity which in principle is readily calculable. This is particularly so for a cavity of rectangular
shape with noncommensurable $a<b<c$ in order to avoid mode degeneracies.

Observe that the walls are made of a conductor and their crystal structure. In turn we expect that several
vacuum fields beside the electromagnetic are going to make contributions to the mass of the walls, but in
the cavity by itself the only contribution can be that of eqn. (15) which is purely electromagnetic. This
suggests a simple experimental procedure in principle.

First, weigh in a precision balance the cavity with its walls in cryogenic conditions, i.e. at $T$ as low as
possible. Then disassemble the cavity and weigh its components. The first measurement under careful
experimental conditions should yield a slightly larger mass than the second one. This excess mass must be
the one associated with the energy $E$ in eqn. (15). This is of course an extremely idealized experiment.

The experiment, if properly accomplished, would give interesting confirmations. First, some confirmation
would be given to the electromagnetic vacuum contribution to inertial mass concept, or quantum vacuum
inertia hypothesis. An important confirmation would be the reality --- or virtuality --- of the zero-point
field. The reality of the zero-point field or quantum vacuum would be unquestionably established. The
difficulty in the experiment is that even though there are many modes inside a cavity resonator, the mass
associated with eqn. (15) should be very small even for a large cavity.

\bigskip\noindent
\centerline{\bf 4. 	THE PHYSICAL BASIS OF THE PRINCIPLE OF EQUIVALENCE}
\bigskip\noindent
We
intend to show not only that the quantum vacuum inertia hypothesis is consistent with GR, but
that it answers an outstanding question regarding a possible physical origin of the force
manifesting as weight [19]. We also intend to show that just as it becomes possible to identify
a physical process underlying the {\bf f}=m{\bf a} postulate of Newtonian
mechanics (as well as its extension to SR [1]), it is possible to identify a
parallel physical process underlying the weak equivalence principle, $m_i=m_g$.

Within the standard theoretical framework of GR and related theories, the equality (or
proportionality) of inertial mass to gravitational mass has to be assumed. It remains
unexplained. As correctly stated by Rindler [20], ``the proportionality of inertial
and gravitational mass for different materials is really a very mysterious fact.''
However here we show that --- at least within the present restriction to
electromagnetism --- the quantum vacuum inertia hypothesis leads naturally and
inevitably to this equality. The interaction between the electromagnetic quantum
vacuum and the electromagnetically-interacting particles constituting any physical
object (quarks and electrons) is identical for the two situations of (a) acceleration
with respect to constant velocity inertial frames or (b) remaining fixed above some
gravitating body with respect to freely-falling local inertial frames.

A related theoretical lacuna involves the origin of the force which manifests itself
as weight. Within GR theory one can only state that deviation from geodesic motion
results in a force which must be an inertia reaction force. We propose that it is
possible in principle to identify a mechanism which
generates such an inertia reaction force, and that in curved spacetime it
acts in the same way as acceleration does in flat spacetime.

Begin by considering a macroscopic, massive, gravitating object, $W$, which is fixed
in space and which for simplicity we assume to be solid, of constant density, and spherical
with a radius
$R$, e.g. a planet-like object. At a distance $r>>R$ from the center of $W$ there is a
small object,
$w$, that for our purposes we may regard as a point-like test particle. A constant
force
$\bf f$ is exerted by an external agent that prevents the small body $w$ from falling
into the gravitational potential of $W$ and thereby maintains $w$ at a fixed point in
space above the surface of $W$. Experience tells us that when the force {\bf f} is
removed, $w$ will instantaneously start to move toward $W$ with an acceleration
{\bf g} and then continue freely falling toward $W$.

Next we consider a freely falling {\it local} inertial frame \Is \ (in the
customary sense given to such a local frame [21]) that is instantaneously at
rest with respect to $w$. 
At $w$ proper time $\t$, that we select to be
$\t=0$, object $w$ is instantaneously at rest at the point \ptg \ of the \Is \
frame. The $x$-axis of that frame goes in the direction from $W$ to $w$ and, since
the frame is freely falling toward $W$, at $\t = 0$ object $w$ appears accelerated
in \Is \ in the $x$-direction and with an acceleration ${\bf g}_w=\hat {x}g$. As
argued below,
$w$ is performing a uniformly-accelerated motion, i.e. a motion with a constant proper
acceleration ${\bf g}_w$ as observed from any neighboring instantaneously comoving
(local) freely-falling inertial frame. In this respect we introduce an infinite collection of local
inertial frames \It, with axes parallel to those of \Is \ and with a common $x$-axis
which is that of \Is . Let $w$ be instantaneously at rest and co-moving
with the frame \It \ at $w$ proper time $\t$.
So the $\t$ parameter representing the $w$ proper time also serves to
parametrize this infinite collection of (local) inertial frames.
Clearly then, \Is \ is the member of the collection with $\t =0$, so that
$I_*=I_{\t=0}$.
At the point in time of coincidence with a given \It,
$w$ is found momentarly at rest at the \ptg \ point of the \It \ frame. We
select also the times in the (local) inertial frames to be $t_{\t}$ and such that
$t_{\t}=0$ at the moment of coincidence when $w$ is instantaneously at rest in \It \
and at the aforementioned \ptg \ point of \It.
Clearly as $I_{\t=0}=I_*$ then $t_*=0$ when $\t=0$. All the frames in the collection
are freely falling toward $W$, and when any one of them is instantaneously at rest
with $w$ it is instantaneously falling with acceleration
${\bf g}=-{\bf g}_w=-\hat{x}g$ with respect to $w$ in the direction of $W$.
It is not difficult to realize that $w$ appears in those frames as uniformly
accelerated and hence performing hyperbolic motion with constant proper
acceleration
${\bf g}_w$.

This situation is equivalent to that of an object $w$ accelerating with respect to
an ensemble of \It \ reference frames in the absence of gravity. In that
situation, the concept of the ensemble of inertial frames,
\It, each with an infinitesimally greater velocity (for the case of positive
acceleration) than the last, and each coinciding instantaneously with an accelerating
$w$ is not difficult to picture. But how does one picture the analogous ensemble for
$w$ held stationary with respect to a gravitating body?

We are free to bring reference frames into existence at will. Imagine bringing a
reference frame into existence at time $\tau=0$ directly adjacent to $w$, but whereas
$w$ is fixed at a specific point above $W$, we let the newly created reference frame
immediately begin free-falling toward $W$. We immediately create a replacement
reference frame directly adjacent to $w$ and let it drop, and so on. The
ensemble of freely-falling local inertial frames bear the same relation to $w$ and to
each other as do the extended \It \ inertial frames used in the case of true
acceleration of
$w$ [1,2].

For convenience we introduce a special frame of reference, $S$, whose $x$-axis
coincides with those of the \It \ frames, including of course \Is, and whose $y$-axis
and
$z$-axis are parallel to those of \It \ and \Is. This frame $S$ stays collocated with 
$w$ which is positioned at the \ptg \ point of the $S$ frame. For \Is \ (and for the
\It \ frames) the frame $S$ appears as accelerated with the uniform acceleration ${\bf
g}_w$ of its point \ptg . We will assume that the frame $S$ is rigid. If so, the
accelerations of points of $S$ sufficiently separated from the $w$ point \ptg \ are
not going to be the same as that of \ptg. This is not a concern however since we will
only need in all frames (\It, \Is \ and $S$) to consider points in a sufficiently
small neighborhood of the \ptg \ point of each frame.

The collection of frames, \It, as well as \Is \ and  $S$, correspond exactly to
the set of frames introduced in [1]. The only differences are, first, that now
they are all local, in the sense that they are only well defined for regions in the
neighborhood of their respective \ptg \ space points; and second, that now \Is \ and
the \It \ frames are all considered to be freely falling toward $W$ and the $S$ frame
is fixed with respect to $W$. Similarly to [1], the $S$ frame may again be
considered to be, relative to the viewpoint of \Is, a {\it Rindler noninertial
frame}. The laboratory frame \Is \ we now call the {\it Einstein laboratory frame},
since now the ``laboratory'' is local and freely falling. We call the collection of
inertial frames \It \ the {\it Boyer family of frames} as he was the first to introduce
them in SED analysis of the Unruh-Davies effect [21].

The relativity principle as formulated by Einstein when proposing SR states that ``all
inertial frames are totally equivalent for the performance of all physical
experiments.''[11] Before applying this principle to the freely-falling frames \Is
\ and \It \ that we have defined above it is necessary to draw a distinction between
these frames and inertial frames that are far away from any gravitating body, such as
$W$. The free-fall trajectories, i.e. the geodesics, in the vicinity of any
gravitating body, $W$, cannot be parallel over any arbitrary distance owing to the
fact that $W$ must be of finite size. 
This means that the principle of relativity can only be applied locally. This was
precisely the limitation that Einstein had to put on his infinitely-extended Lorentz
inertial frames of SR when starting to construct GR [11,20,22].

We adopt the principle of local Lorentz invariance (LLI) which can be stated,
following Will [11], as ``the outcome of a local nongravitational test experiment is
independent of the velocity of the freely falling apparatus.'' A non-gravitational
test experiment is one for which self-gravitating effects can be neglected. We also
adopt the assumption of space and time uniformity, which we call the uniformity
assumption (UA) and which states that the laws of physics are the same at any time or
place within the Universe. Again, following Will [11] this can be stated as ``the
outcome of any local nongravitational test experiment is independent of where and
when in the universe it is performed.'' We do not concern ourselves with physical or
cosmological theories that in one way or another violate UA, e.g. because they involve
spatial or temporal changes in fundamental constants [23].

Locally, the freely falling local Lorentz frames which we now designate with a
subscript
$L$ --- i.e. \Isl \ and \Itl \ ---
are entirely equivalent to the \Is \ and \It \ extended frames of [1]. The
free-falling Lorentz frame \Itl \ locally is exactly the same as the extended \It.
Invoking the LLI principle we can then immediately conclude that the
electromagnetic zero-point field, or electromagnetic quantum vacuum, that can be
associated with
\Itl \ must be the same as that associated with  \It. From the viewpoint of the
local Lorentz frames \Isl \ and \Itl \ the body $w$ is undergoing uniform
acceleration and therefore for the same reasons as presented in [1] a peculiar
acceleration-dependent force arises, that for concreteness we shall call a drag force.

{\it These formal arguments demonstrate that the analyses of [1] which found
the existence of a Rindler frame force in an accelerating reference frame translate and
correspond exactly to a reference frame fixed above a gravitating body.}
In the same manner that light rays are deviated from straight-line propagation by a
massive gravitating body $W$, the other forms of electromagnetic radiation, including
the electromagnetic zero-point field rays (in the SED approximation), are also
deviated from straight-line propagation. Not surprisingly this creates an anisotropy
in the otherwise isotropic electromagnetic quantum vacuum.

In [1] we interpreted the drag force exerted by the electromagnetic quantum vacuum radiation as
the inertia reaction force of an object that is being forced to accelerate through the
electromagnetic quantum vacuum field. Accordingly, in the present situation, the
associated nonrelativistic form of the inertia reaction force should be

$$\Fszp=-m_i {\bf g}_w \eqno(22)
$$
where ${\bf g}_w$ is the acceleration with which $w$ appears in the local inertial
frame \Is. As shown in [1] the coefficient $m_i$ is

$$m_i= \left[ {V_0 \over c^2} \int \ew \A d\w \right] 
\eqno(23)$$
where $V_0$ is the proper volume of the object, $c$ is the speed of light, $\hbar$ is
Planck's constant divided by $2 \pi$ and $\ew$, where $0 \le \ew \le 1$, is a
function that spectralwise represents the relative strength of the interaction between
the zero-point field and the massive object, interaction which acts
to oppose the acceleration. If the object is just a single particle, the spectral
profile of
$\ew$ will characterize the electromagnetically-interacting particle. It can also
characterize a much more extended object, i.e. a macroscopic object, but then the
$\ew$ will have much more structure (in frequency). We should expect different shapes
for the electron, a given quark, a composite particle like the proton, a molecule, a
homogeneous dust grain or a homogeneous macroscopic body. In the last case the $\ew$
becomes a complicated spectral opacity function that must extend to extremely high
frequencies such as those characterizing the Compton frequency of the electron and
even beyond.

Now, however, what appears as inertial
mass,
$m_i$, to the observer in the local \Isl \ frame is of course what corresponds to passive
gravitational mass, $m_g$, and it must therefore be the case that

$$m_g= \left[ {V_0 \over c^2} \int \ew \A d\w \right] .
\eqno(24)$$
As done in [1], Appendix B, it can be shown that the right hand side indeed
represents the energy of the electromagnetic quantum vacuum enclosed within the
object's volume and able to interact with the object as manifested by the $\ew$
coupling function. A more thorough, fully covariant development can also be
implemented to show that the force expression of eqn. (1) can be extended to the
relativistic form of the inertia reaction force as in [1], Appendix D. (This
development also served to obtain the final form of $m_i$ given above in eqn. (3)
eliminating a spurious 4/3 factor.)
\footnote{$^2$}{We use this opportunity to correct a minor transcription error that
appeared in the printed version of Ref. [1]. In Appendix D, p. 1100, the minus sign in
eqn. (D8) is wrong. It should read
$$\eta^{\nu} \eta_{\nu}=1 \eqno({\rm D}8)
$$
and the corresponding signature signs in the line just above eqn. (D8) are the
opposite of what was written and should instead read $(+ - - \ - )$.
}
Summarizing what we have shown in this section is that if a force {\bf f} is applied
to the $w$ body just large enough to prevent it from falling toward the body $W$,
then in the non-relativistic case that force is given by

$${\bf f}=-m {\bf g} \eqno(25)
$$
where we have dropped the nonessential subscripts $i$ and $g$ and superscripts,
because it is now clear that $m_i=m_g=m$ follows from the quantum vacuum inertia
hypothesis.

The physical basis for the principle of equivalence is the fact that accelerating
through the electromagnetic quantum vacuum is identical to remaining fixed in a
gravitational field and having the electromagnetic quantum vacuum fall past on curved
geodesics. In both situations the observer will experience an asymmetry in the radiation
pattern of the electromagnetic quantum vacuum which results in a force --- either the inertia
reaction force or weight --- which become then within this more general Einsteinian perspective the same thing.

\bigskip\noindent
\centerline{\bf 5. CONSISTENCY WITH EINSTEIN'S GENERAL RELATIVITY}
\bigskip\noindent
The statement that $m_i=m_g$ constitutes the weak equivalence principle (WEP). Its
origin goes back to Galileo and Newton, but it now appears, as shown in the previous
section, that this principle is a natural consequence of the quantum vacuum inertia
hypothesis. The strong equivalence principle (SEP) of Einstein consists of the WEP
together with LLI and the UA. Since the quantum vacuum inertia hypothesis and its
extension to gravity allow us to obtain the WEP assuming LLI and the UA, this
approach is consistent with all theories that are derived from the SEP. In addition
to GR, the Brans-Dicke theory is derived from SEP as are other
lesser known theories [11], all distinguished from each other by
various particular assumptions.

All theories that assume the SEP are called {\it metric theories}. They are
characterized by the fact that they contemplate a bending of spacetime associated
with the presence of matter. Two important consequences of the LLI-WEP-UA combination
are that light bends in the presence of matter and that there is a gravitational
Doppler shift. Since the quantum vacuum inertia hypothesis is consistent with this
same combination, it would also require that light bends in the presence of
gravitational fields. This can, of course, be interpreted as a change in spacetime
geometry, the standard interpretation of GR.
However this does not mean that we have explained the mechanism for the actual bending of space-time in the
vicinity of a material object. This is the origin of so-called active gravitational mass that still
requires an explanation within the viewpoint of the quantum vacuum inertia hypothesis.

\bigskip\noindent
\centerline{\bf 6. A RECENT ALTERNATIVE VACUUM APPROACH TO GRAVITY}
\bigskip\noindent
The idea that the vacuum is ultimately responsible for gravitation is not new. It
goes back to a proposal of Sakharov [24] based on the work of Zeldovich [25] in
which a connection is drawn between Hilbert-Einstein action and the quantum vacuum.
This leads to a view of gravity as ``a metric elasticity of space'' (see Misner,
Thorne and Wheeler [26] for a succinct review of this concept). Following Sakharov's
idea and using the techniques of SED, Puthoff proposed that gravity could be
construed as a form of van der Waals force [27]. Although interesting and
stimulating in some respects, Puthoff's attempt to derive a Newtonian inverse square
force of gravity proves to be unsuccessful [28,29,30,31].

An alternative approach has recently been developed by Puthoff [32] that is based on
earlier work of Dicke [33] and of Wilson [34]: a polarizable vacuum model of
gravitation. In this representation, gravitation comes from an effect by massive
bodies on both the permittivity, $\epsilon_0$, and the permeability, $\mu_0$, of the
vacuum and thus on the velocity of light in the presence of matter.

That is clearly an alternative theory to GR since it does not involve
curvature of spacetime. On the other hand, since spacetime curvature
is by definition inferred from light propagation in relativity theory, the polarizable
vacuum gravitation model may be labelled a pseudo-metric theory of gravitation since
the effect of variation in the dielectric properties of the vacuum by massive objects
on light propagation are approximately equivalent to GR spacetime curvature as long
as the fields are sufficiently weak. In the weak-field limit, the polarizable vacuum
model of gravitation duplicates the results of GR, including the classic tests
(gravitational redshift, bending of light near the Sun, advance of the perihelion of
Mercury). Differences appear in the strong-field regime, which should lead to
interesting tests.

\bigskip\noindent
\centerline{\bf 7. DERIVATION OF NEWTON'S LAW OF GRAVITATION}
\bigskip\noindent
Our approach allows us to derive a Newtonian form of gravitation in the weak-field
limit based on the quantum vacuum inertia hypothesis, local Lorentz invariance and
geometrical considerations. In [1] we showed how an asymmetry that appears in
the radiation pattern of the electromagnetic quantum vacuum, when viewed from an
accelerating reference frame, leads to the appearance of an apparent non-zero
momentum flux. Individual and collective interaction between the
electromagnetically-interacting particles (quarks and electrons) comprising a
material object and the electromagnetic quantum vacuum generates a reaction force that may be
identified with the electromagnetic contribution to the inertia reaction force as it has the right form for
all velocity regimes. In particular in the low velocity limit it is exactly proportional to the acceleration
of the object.

In $\S 4$ we showed from formal arguments based on the principle of local Lorentz
invariance (LLI) that an exactly equivalent force must originate when an object is
effectively accelerated with respect to free-falling Lorentz frames by virtue of
being held fixed above a gravitating body, $W$. We infer that the presence of a
gravitating body must distort the electromagnetic quantum vacuum in exactly the same
way at any given point as would the process of acceleration such that ${\bf a}=-{\bf
g}$. Simple geometrical arguments [35] now suffice to show that the gravitational
force can only be the Newton inverse-square law with distance (in the weak field
limit).

It has been shown above that (outside $W$) the {\bf g} field of eqn. (25) generated
by
$W$ is central, i.e., centrally distributed with spherical symmetry around
$W$. It has to be radial, with its vectorial direction parallel to the corresponding
radius vector {\bf r} originating at the center of mass of $W$ where we locate the
origin of coordinates. The spherical symmetry implies that {\bf g} is radial in the
direction
$-\hat{\bf r}$, and depends only on the $r$-coordinate, ${\bf g} = - \hat{\bf r} g(r)$.

The field is clearly generated by mass. A simple symmetry argument based on Newton's Third Law,
here omitted for brevity, shows that indeed if we scale the mass
$M$  of
$W$ by a factor $\alpha$ to become $\alpha M$ then the resulting {\bf g} must be of the
form $\alpha {\bf g}$. If $M$ goes to zero, {\bf g} disappears.
The mass $M$ must be the source of field
lines of {\bf g}, and these field lines can be discontinuous only where mass is
present. The field lines can be neither generated nor destroyed in free space.  Since
{\bf g} is the force on a unit mass, we must expect that {\bf g} behaves as a vector,
and specifically that {\bf g} follows the laws of vector addition. Namely, if two
masses
$M_1$ and $M_2$ in the vicinity of each other generate fields ${\bf g}_1$ and ${\bf
g}_2$ respectively, the resulting {\bf g} at any given point in space should be the
linear superposition of the ${\bf g}_1$ and ${\bf g}_2$ fields at that given point, namely their vector sum

$${\bf g}={\bf g}_1+{\bf g}_2 .
\eqno(26)
$$
Finally, from the argument that leads to eqn. (25) we can see
that the field {\bf g} must be unbounded, extending essentially to infinity.

With all of these considerations, clearly the lines of {\bf g} must obey the
continuity property outside $W$. If there is no mass present inside a volume, $V$,
enclosed by a surface $S$, we expect, using Gauss' divergence theorem, that

$$0=\oint_{S(V)} {\bf g} \cdot {\bf n} \ dS = \int_V \nabla \cdot {\bf g} \ dV ,
\eqno(27)
$$
but since $V$ is arbitrary this tells us that outside the massive body $W$

$$\nabla \cdot {\bf g}=0 , \ \ \ \ r > R \ . \eqno(28)
$$
In the presence of our single, spherically symmetric massive body, $W$, but outside that body,
{\bf g} is radial with respect to the center of $W$. Hence we must have ${\bf g} = \hat{r} \Phi(r)$, and
therefore from eqn. (28), because $\Phi$ is only a function of $r$, it necessarily follows that {\bf g} is
of the form

$${\bf g} \sim -\hat{\bf r} {1 \over r^2} , \eqno(29)
$$
where the restriction $r>R$ is hereafter understood.
Since the field {\bf g} is also proportional to the mass $M$ which is its origin, we
conclude that {\bf g} must be of the form

$${\bf g} = -\hat{\bf r} {GM \over r^2} , \eqno(30)
$$
where $G$ is a proportionality constant and from eqn. (17) we have that

$${\bf F} = -\hat{\bf r} {GMm \over r^2} , \eqno(31)
$$
which is Newton's law of gravitation. It is remarkable that after finding
the central and radial character of {\bf g} by means of the vacuum approach of
our quantum vacuum inertia hypothesis, one can immediately obtain Newtonian
gravitation, an endeavour keenly but unsuccessfully pursued for quite some time from
the viewpoint of the vacuum fields [24,25,26] and in particular of SED
[27,28,29,30,31]. In this last case (SED), it was proposed that gravity was a force of
the van der Waals form, a view which has been shown to be unsuccessful [31].

As a final point here we observe that when the two bodies of masses $M$ and $m$ are reduced to point
particles, a simple argument based on the symmetry of the situation shows that the two roles of the masses
$m$ and
$M$ (passive and active) can be interchanged and that similarly the force on $M$ due to $m$ should be equal
in magnitude but opposite in direction to that of eqn. (22) so that Newton's third law of mechanics is
satisfied.

\bigskip\noindent
\centerline{\bf 8. ON THE ORIGIN OF WEIGHT}
\bigskip\noindent
We have established that the
quantum vacuum inertia hypothesis leads to a force in eqn. (22) which adequately
explains the origin of weight in a Newtonian view of gravitation. How is this
consistent with the geometrodynamic view of GR? Geometrodynamics specifies the
effect of matter and energy on an assumed pliable spacetime metric. That defines the
geodesics which light rays and freely-falling objects will follow. However there is
nothing in geometrodynamics that points to the origin of the inertia reaction
force when geodesic motion is prevented, which manifests in special circumstances as
weight.

Geometrodynamics merely assumes that deviation from geodesic motion results in
inertial forces. That is, in fact, true, but as it stands it is devoid of any
physical insight. What we have shown above is that an
identical asymmetry in the quantum vacuum radiation pattern will arise due to either
true acceleration or to effects on light propagation by the presence of gravitating
matter. In the case of
true acceleration, the resulting force is the inertia reaction force. In the case of
being held stationary in a non-Minkowski metric, the resulting force is the weight,
which is also the enforcer of geodesic motion for freely-falling objects.

\bigskip\noindent
\centerline{\bf 9. A PHANTOM ANOMALY}
\bigskip\noindent
Following the reasoning of $\S 7$ one would conclude that at every point of fixed $r$
above $W$ there is an inflowing energy-momentum flux from the falling electromagnetic quantum
vacuum. This would seem to imply a continuous energy flux toward and into $W$, an apparently
paradoxical situation consisting of quantum vacuum-originating energy streaming toward $W$
from all directions in space. But this paradox is only apparent, as we show.

It is true that for a freely falling observer attached to a local inertial frame \Itl \  which
is falling past the surface of the Earth and towards its center ($\S 4$), the object $w$, that
we take now to be located on the surface of the Earth, appears to be uniformly accelerated
with constant proper acceleration $\g_w = - \g$. Thus this observer concludes as he freely
falls that the mass of $w$ must grow as $\gamma_{\tau} m$, where $m$ is the rest mass of $w$
and $\gamma_{\tau}$ is the Lorentz factor, $(1-\beta_{\tau}^2 )^{-1/2}$.

Concommitantly, this observer of \Itl , in light of our discussion in App. B of Ref. [1] and
because of the SEP, asserts that the energy content of $w$, which for simplicity we take here
as purely electromagnetic and coming from the electromagnetic vacuum, is steadily increasing.
Eqn. (33) of Ref. [1] gives an energy growth rate $dE/dt = {\bf f} \cdot {\bf v}$, where {\bf
f} in the present case corresponds to the force of support against the Earth's gravity exerted
on $w$ by the surface of the Earth. Clearly then this freely falling observer of \It \
concludes that objects on the surface of the Earth are steadily increasing their internal
energy at the expense of the falling vacuum, namely of the electromagnetic zero-point field of
\It , that freely falls with \It .

However when exactly the same consideration is made but now from the point of view of an
observer fixed on the surface of the Earth and in the neighborhood of $w$, the conclusion is
quite different. In this case the observer is attached to $S$, the local Rindler non-inertial
frame where $w$ is fixed and that coincides with the surface of the Earth around the spot where
$w$ is located. To calculate $dE/dt$ as seen from $S$ we use the local instantaneously
comoving, freely-falling inertial frame \Isl = \Itzl \ which at that point in time co-moves with
$w$. In that frame \Isl \ the particle velocity is of course zero, and therefore the rate
of energy growth is $dE/dt=0$. So the observer in $S$ concludes that $w$ does not change its energy
content.

The above shows that for an observer on the surface of the Earth there is no real transfer of
energy from the vacuum to objects on the surface of the Earth. In this respect the energy
growth above, that a freely falling observer sees as time progresses, is purely a kinematical
effect that comes form the fact that in the freely-falling frame the velocity of the object $w$
is growing with time and does not vanish except at the instant of coincidence of \Isl \ with $S$,
at $\tau=0$. In special relativity, a constantly accelerating spaceship is said by an
external observer to be increasing its mass steadily as
$\gamma(\tau) m$, but of course an observer onboard that craft will not perceive any
relativistic mass increase.

On the other hand as seen in $S$, for the frame at the surface of the Earth, even though
$dE/dt$ vanishes, the force {\bf f} does not vanish, as shown in Ref. [1]. The force {\bf f} is needed to
compensate the force ${\bf f}_{zp}$, i.e. ${\bf f}_{zp}=-{\bf f}$, exerted on
$w$ by the freely-falling vacuum of \It , that indeed corresponds to what we called above the
Rindler frame force.

\bigskip\noindent
\centerline{\bf 10. DISCUSSION}
\bigskip\noindent
In light of what has been proposed herein, what can we
elucidate about the nature of the gravitational field? In the low fields
and low velocities version, or the Newtonian limit, we have seen above that gravity
manifests itself as the attractive force per unit mass, {\bf g}, of eqn. (30) that
pulls any massive test body present at a given point in space towards the body $W$
that originates the field. Since we assumed the Einstein LLI principle and from this
derived the WEP, this, together with the very natural UA of invariance in the laws of
physics throughout universal spacetime, leads us to the Einstein SEP which necessarily
implies the spacetime bending representation of the generalized gravitational
field [11,20].

A simple thought experiment (Einstein's lift) immediately shows that light rays
propagate along geodesics, and more specifically along null geodesics [20,26,36]. The
spacetime bending is dramatically evident when a light ray goes from one side to the
other of the freely falling elevator. For the observer attached to the elevator's
frame that indeed acts as a LLI frame, the light ray propagates in a straight line
from one side to the other of the elevator. But for the stationary observer who sees
the elevator falling with acceleration {\bf g}, the light ray bends along a path that
locally is seen as a parabolic curve. Undoubtedly the most natural explanation for
the stationary observer is that spacetime bends and therefore the association that
this bending is a manifestation of the gravitational field of $W$, or rather that
this bending of spacetime is the gravitational field itself [36].   

Starting from the above fact taken as a given, the various metric theories proceed
from there to formulate their equations. In the version of Brans and Dicke a
scalar-tensor field is assumed. In Einstein's GR only a tensor
field is proposed. Following this maximally simplistic proposal and guided by general
considerations of general covariance (the need of arbitrary coordinates and tensor
laws) Einstein was led to his field equations in the presence of matter\dots and then GR
naturally unfolded [11,20,26,36].

We have shown here that our inertia proposal of [1] leads us, when limited by
the LLI principle, to the metric theories and therefore that it is consistent with
those theories and in particular with Einstein's GR. In addition, there is the
following interesting feature of our proposal.

From our analysis in [1], and in particular in Appendix B of [1], it was
made clear that within the quantum vacuum inertia hypothesis proposed therein, the mass of the
object,
$m$, could be viewed as the energy in the equivalent electromagnetic quantum vacuum field
captured within the structure of the object and that readily interacts with the object. This
view properly and accurately matched with the complementary view,
exposed in other parts of the paper [1], that presented inertia as the result of a
vacuum reaction effect, a kind of drag force exerted by the vacuum field on
accelerated objects. Quantitatively, both approaches lead to exactly the same inertial
mass and moreover they are partly complementary. Both viewpoints are needed. One
could not exist without the other. They were the two sides of the same coin.

The question is now why massive objects, when freely falling, also follow geodesic
paths. The tempting view suggested here is that, as massive bodies have a mass that is made of the vacuum
electromagnetic energy contained within their structure and that readily interacts
with such structure (according to
our analysis of [1], Appendix B), it is no surprise that geodesics are their natural path of
motion during free fall. Electromagnetic radiation has been shown by Einstein to
follow precisely geodesic paths. The only difference now is that, as the radiation
stays within the accelerated body structure and is contained within that structure
and thereby its energy center moves subrelativistically, these geodesics are just
time-like and not null ones as in the case of freely propagating light rays.

We illustrate this with an example. Imagine a freely-falling electromagnetic cavity
with perfectly reflecting walls of negligible weight, so that all the weight is due
to the enclosed radiation. A simple plane wave mode decomposition shows that although
individual wavetrains do still move at the speed of light, the center of energy of
the radiation inside the cavity moves subrelativistically as the wavetrains reflect
back and forth. The wavetrains do indeed propagate along null geodesics, but the
center of energy propagates only along a time-like geodesic.

Neither our approach nor the conventional presentations of GR for
that matter, can offer a physical explanation of the mechanism of the bending of
spacetime as related to energy density. Misner, Thorne and Wheeler [26] present six
different proposed explanations. The sixth is the one we already mentioned due
to Sakharov [24,25] which starts from general vacuum considerations. As our approach
starts also from vacuum considerations, it naturally fits better the concept of the
conjecture of Sakharov [24] and Zeldovich [25] than the other proposals but it is not
inconsistent with any of them. In particular the strictly formal proposal of Hilbert
[26] that introduces the so-called Einstein-Hilbert Action, is also at the origin of
the Sakharov proposal. We plan to devote more work to exploring the connection of our
inertia [1] and gravity approach to the approach proposed by Sakharov [24]. This we
leave for a future publication.

\bigskip\noindent
\centerline{\bf 11. CONCLUSIONS}
\bigskip\noindent
The principal conclusions of this paper are:

(1) {\it Identity of inertial mass with gravitational mass, $m_i=m_g$ naturally follows from the quantum
vacuum inertia hypthesis.} It has been shown that the approach of [1] contains this peculiar feature that,
so far and as we know, has never been explained. Rindler [20] calls this feature ``a very
mysterious fact,'' as indeed it has been up to the present. We expect with this work
to have shed some light on this peculiar feature.

(2) {\it The quantum vacuum inertia hypothesis is consistent with Einstein's GR.} We
have already commented above, in particular in $\S 8$, on this interesting feature
presented in $\S 5$ that puts the vacuum inertia approach of [1] within the
mainstream thought of contemporary gravitational theories, specifically within that
of theories of the metric type and in particular in agreement with GR.

(3) {\it Newton's gravitational law naturally follows from the quantum vacuum inertia hypothesis.}
By means of a simple argument based on potential theory we show how to obtain in a
natural way Newton's inverse square force with distance from our vacuum approach to
inertia of [1]. The simplicity of our approach contrasts with previous attempts to
accomplish this within the framework of SED theory.

(4) {\it An origin of weight and a physical mechanism to enforce motion along geodesic
trajectories for freely-falling objects is contained in the hypothesis}.
We have 
shown how this approach to inertia answers a fundamental question left open
within GR, viz. is there a physical mechanism that generates the inertia reaction force
when non-geodesic motion is imposed on an object and which can manifest specifically
as weight. Or put another way, while geometrodynamics dictates the spacetime metric
and thus specifies geodesics, is there an identifiable mechanism for enforcing motion
along geodesic trajectories? The quantum vacuum inertia hypothesis represents a
significant first step in providing such a mechanism.

(5) {\it A physical basis is provided for the relativistic mass increase}. In the approach of the quantum
vacuum inertia hypothesis, it becomes physically evident how the Lorentz factor
$\gamma_{\tau}$, which characterizes a space-time geometry relationship, acquires the physical role of
relativistic mass increase parameter. It also becomes clear why the relativistic mass increase must become
infinite at
$c$: One cannot propagate the effect of any forces originating in zero-point field effects at speeds faster
than
$c$.

(6) {\it The interior (excluding the walls) of a cavity resonator is presented as the archetype of a system
wherein only the electromagnetic vacuum contributes to the mass.} This allows an expression for a resulting
inertial mass with no free parameters. This mass in principle can be measured.

(7) {\it An experimental prediction has been made that the mass of the resonant electromagnetic zero-point
field modes within a cavity should add to the mass of the cavity structure.}

\bigskip\noindent
\centerline{\bf ACKNOWLEDGEMENTS}
\bigskip\noindent
AR received partial support from the California Institute for Physics and
Astrophysics via a grant to Cal. State Univ. at Long Beach and via release time from the Office of Academic
Affairs, CSULB. This work was partially funded  under NASA contract NASW-5050.

{

\bigskip

\parskip=0pt plus 2pt minus 1pt\leftskip=0.25in\parindent=-.25in 

\centerline{\bf REFERENCES}
\bigskip

[1] A. Rueda and B. Haisch, Found. Phys. {\bf 28} (1998) 1057; see also A. Rueda and B. Haisch, Phys.
Lett. A {\bf 240} (1998) 115

[2] B. Haisch, A. Rueda, Y. Dobyns, Ann. Physik (Leipzig) {\bf 10} (2001) 393

[3] B. Haisch, A. Rueda  and H.E. Puthoff, Phys. Rev A {\bf 48}
(1994) 678.

[4] R. Loudon, The Quantum Theory of Light (2nd ed.), Clarendon Press, Oxford (1983), p. 5.

[5] H. Sunahata and A. Rueda, in prep. (2005)

[6] P.J.E. Peebles and B. Ratra, Rev. Mod. Phys. {\bf 75} (2003) 559 (in particular $\S$IIB)

[7] B. Haisch and A. Rueda, Phys. Lett. A {\bf 268} (2000) 224

[8] T. H. Boyer, Phys. Rev. D {\bf 11} (1975) 790

[9] P. W. Milonni, The
Quantum Vacuum: An Introduction to Quantum Electrodynamics, $\S 8.2$, Academic Press (1994)

[10] L. de La Pe\~na and A.M. Cetto, The
Quantum Dice -Ð An Introduction to Stochastic Electrodynamics, Kluwer Acad. Publ.,
Fundamental Theories of Physics Series, Dordrecht, Holland (1996) and references
therein.

[11] C.W. Will, Theory and Experiment in Gravitational Physics, Cambridge University
Press, Cambridge (1993) pp 22--24

[12] H. Weyl, J. Math, {\bf 143} (1913) 177; H. Weyl, R. Cir. Mat. Palermo {\bf 39} (1913) 1

[13] M. Ibison and B. Haisch, Phys. Rev. A {\bf 54} (1996) 2737

[14] J. D. Jackson, Classical Electrodynamics (3rd ed.), Wiley and Sons, New York (1999)
p. 558

[15] A. Yariv, Quantum Electronics, Wiley, New York, (1967)

[16] B. Saleh and M. C. Teich, Fundamentals of Photonics, Wiley, New York (1991)

[17] E. A. Hinds, ``Perturbative Cavity Quantum Electrodynamics'' in Cavity Quantum Electrodynamics, P. R.
Berman (ed.), pp. 1--54, Academic Press, Inc., New York (1994)

[18] W. H. Louisell, Quantum Statistical Properties of Radiation, Wiley, New York (1975)

[19] Y. Dobyns, A. Rueda and B. Haisch, Found. Phys. {\bf 30} (1) (2000) 59

[20]  W. Rindler, Essential Relativity Ð Special, General
and Cosmological, Springer Verlag, Heidelberg (1977), p. 17.

[21] T. H. Boyer, Phys. Rev. D {\bf 21} (1980) 2137 and Phys. Rev. D {\bf 29} (1984) 1089

[22] A. Einstein, Ann. Phys. {\bf 35} (1911) 898. For a translation see C.W. Kilmister,
General Theory of Relativity, Pergamon, Oxford (1973), pp. 129--139. 

[23] See,e.g., P.A.M. Dirac, Directions in Physics, Wiley, New York (1978), in
particular Section 5,  ``Cosmology and the gravitational constant,'' pg. 71 ff.
 
[24] A.D. Sakharov, Doklady Akad. Nauk S.S.S.R. {\bf 177} (1967) 70 (English translation
in Sov. Phys. Doklady {\bf 12} (1968) 1040

[25] Yu. B. Zeldovich. Zh. Eksp. \& Teor.
Fiz. Pis'ma {\bf 6} (1967) 883 (English translation in Sov. Phys. -- JETP Lett. {\bf 6} (1967)
316

[26] C.W. Misner, K.S. Thorne and J.A. Wheeler, Gravitation, Freeman,
New York, (1971) pp 426--428.

[27] H.E. Puthoff, Phys. Rev. A {\bf 39} (1989) 2333

[28] S. Carlip, Phys. Rev A {\bf 47} (1993) 3452

[29] H.E. Puthoff, Phys. Rev. A {\bf 47} (1993) 3454

[30] B. Haisch, A. Rueda and H.E. Puthoff, Spec. Science and Technology {\bf 20} (1997) 99

[31] D. C. Cole, A. Rueda and K. Danley, Phys. Rev. A, {\bf 63} (2001) O54101-1, -2

[32] H.E. Puthoff, Foundations Phys. {\bf 32} (2002) 927

[33] R.H. Dicke,
``Gravitation without a principle of equivalence,'' Rev. Mod. Phys. {\bf 29} (1957) 363; see also R.H.
Dicke, ``Mach's Principle and Equivalence,'' in Proceedings of the International School of Physics ``Enrico
Fermi'' Course XX, Evidence for Gravitational Theories, ed. C. Moller, Acad. Press, New York, (1961), pp
1--49.

[34] H.A. Wilson, Phys. Rev. {\bf 17}, (1921) 54

[35] Our arguments will partially be based on potential theory, see, e.g. O.D.
Kellog, Foundations of Potential Theory, Dover, New York  (1953) pp 34-39, in
particular see Ex 3, p. 37.

[36] See. e.g., R.M. Wald, General Relativity, Univ. of Chicago Press, Chicago, (1984)
pg. 67; and for a more popularizing account, R.M. Wald, Space, Time and Gravity, Second
Edition, Univ. of Chicago Press, Chicago (1992) Ch. 3 and in particular pp 33--34.

}

\bye